\pdfoutput=1
\documentclass[aps,twocolumn,superscriptaddress]{revtex4-1}
\usepackage{graphicx}
\usepackage{amsmath}
\usepackage{amssymb}

\def\ben{\begin{equation}}
\def\een{\end{equation}}
\def\dfts{DFT$+\Sigma$ }
\def\mh{\mathbf{H}}
\def\mv{\mathbf{V}}
\def\mg{\mathbf{\Gamma}}
\def\tmg{\tilde {\mathbf{\Gamma}}}
\def\mgf{\mathbf{G}}
\def\ms{\mathbf{S}}
\def\msig{\mathbf{\Sigma}}
\def\tmsig{\tilde {\mathbf{\Sigma}}}

\begin{document}

\title{Energy-dependent resonance broadening in symmetric and asymmetric molecular junctions from an \emph{ab initio} non-equilibrium Green's function approach}
\author{Zhen-Fei Liu}
\affiliation{Molecular Foundry and Materials Sciences Division, Lawrence Berkeley National Laboratory, Berkeley, California 94720, USA}
\author{Jeffrey B. Neaton}
\affiliation{Molecular Foundry and Materials Sciences Division, Lawrence Berkeley National Laboratory, Berkeley, California 94720, USA}
\affiliation{Department of Physics, University of California, Berkeley, California 94720, USA}
\affiliation{Kavli Energy Nanosciences Institute at Berkeley, Berkeley, California 94720, USA}
\date{\today}

\begin{abstract}
The electronic structure of organic-inorganic interfaces often feature resonances originating from discrete molecular orbitals coupled to continuum lead states. An example are molecular junctions, individual molecules bridging electrodes, where the shape and peak energy of such resonances dictate junction conductance, thermopower, I-V characteristics and related transport properties. In molecular junctions where off-resonance coherent tunneling dominates transport, resonance peaks in the transmission function are often assumed to be Lorentzian functions with an energy-independent broadening parameter $\Gamma$. Here we define a new energy-dependent resonance broadening function, $\Gamma(E)$, based on diagonalization of non-Hermitian matrices, which can describe resonances of a more complex, non-Lorentzian nature and can be decomposed into components associated with the left and right lead, respectively. We compute this quantity via an \emph{ab initio} non-equilibrium Green's function approach based on density functional theory for both symmetric and asymmetric molecular junctions, and show that our definition of $\Gamma(E)$, when combined with Breit-Wigner formula, reproduces the transmission calculated from DFT-NEGF. Through a series of examples, we illustrate how this approach can shed new light on experiments and understanding of junction transport properties in terms of molecular orbitals.
\end{abstract}

\maketitle

In quantum mechanics, resonances can arise when discrete states are coupled to a continuum\cite{F61}. Resonances are prevalent in many fields of physics and chemistry, such as autoionization\cite{F61}, negative ions\cite{BC94}, electron-molecule scattering\cite{T72}, and molecular junctions\cite{Datta05,Nitzan}. Formally, the continuum can be considered to introduce a complex self-energy to the discrete states, including a real part that corresponds to a shift in energy of the discrete states, and an imaginary part that is associated with the resonance broadening, line width, or equivalently lifetime of the resonance states. Several theories, building on non-Hermitian quantum mechanics\cite{M11}, have been developed to characterize resonance states and their broadening, such as complex scaling\cite{M98} and its generalization to density functional theory\cite{WW10,ZE12}; complex absorbing potentials\cite{MPNE04}; complex coordinates and basis functions\cite{MR78}; the stabilization method\cite{HT70} and its modified version\cite{RMMT94}; and projection operators\cite{F58}, to name a few. Here, we add to this body of work in the context of transport in molecular junctions.

When molecules are adsorbed on surfaces or are bridging electrodes in molecular junctions, their discrete orbitals are coupled to continuum states, which give rise to molecular resonances\cite{Datta05,Nitzan}. In the case of molecular junctions, the shape and peak energies of these resonance states are recognized as the peaks in the transmission function $T(E)$. In an off-resonance coherent transport regime, they determine the transmission at the junction Fermi energy, $T(E_F)$, which is directly proportional to the linear-response conductance. The peak height is determined by the symmetry of the junction, and the shape is determined by the coupling of molecular orbitals to the electrodes. The peak position relative to $E_F$ is determined by factors discussed elsewhere\cite{NHL06,QVCL07,LWYA14}. In the so-called  ``wide band limit'', the broadening is energy-independent and the lineshape takes up a Lorentizian form\cite{Datta05,Nitzan}. However in general, the wide band limit does not apply, and a simple broadening parameter does not suffice. Further, when a molecule is asymmetrically coupled to two or more different leads, it is challenging to separate the contributions of the leads to the broadening. In this work, we first briefly review the Lorentizian model and the wide band limit, and then discuss its limitations in more general cases. After that, we develop a method to restore the Lorentizian model, but with an energy-dependent broadening function, based on diagonalization of non-Hermitian matrices within a non-equilibrium Green's function (NEGF) formalism\cite{transiesta}. The NEGF formalism is a natural choice for studies of charge transport through molecular junctions. The non-Hermiticity of the matrices arises from complex self-energies, which are at the origin of resonance widths.

Consider a single discrete state with energy $\epsilon_s$ coupled to continuum states $\{\epsilon_l\}$ and $\{\epsilon_r\}$ in the left and right leads, respectively. Based on the Landauer formula\cite{L57}, the energy-dependent transmission coefficient through the single level can be written as
\ben
T(E)=\frac{\Gamma_L(E)\Gamma_R(E)}{\left[E-\epsilon_s-\mbox{Re}\Sigma(E)\right]^2+\left[\mbox{Im}\Sigma(E)\right]^2},
\label{lore}
\een
where $\Gamma_L(E)=-2\mbox{Im}\Sigma_L(E)=2\pi\int|V_{sl}|^2\delta(E-\epsilon_l)\rho(\epsilon_l)\, d\epsilon_l$ [and similarly for $\Gamma_R(E)$]. $V_{sl}$ is the coupling between the single discrete state and left continuum, $\rho(\epsilon_l)$ is density of states of the left continuum, and $\Sigma=\Sigma_L+\Sigma_R$ is the self-energy due to the two baths. For completeness, $\mbox{Re}\Sigma_L(E)={\cal P}\int|V_{sl}|^2\rho(\epsilon_l)/(E-\epsilon_l)\, d\epsilon_l$, where ${\cal P}$ is principal part of the integral.

Eq. \eqref{lore}, with the energy-dependent $\Gamma(E)$'s, is exact, for $\emph{one}$ discrete level coupled to two baths. In the case that $\rho(\epsilon_l)$ does not vary appreciably with energy, the so-called ``wide band limit'' holds, where $\gamma=-2\mbox{Im}\Sigma(E)$ is a constant, for \emph{both} left and right leads\cite{Nitzan}, neglecting any energetic shift to the discrete state. The transmission then reads
\ben
T(E)=\frac{\gamma_L\gamma_R}{(E-\epsilon_s)^2+(\gamma_L+\gamma_R)^2/4},
\label{wbl}
\een
which is the well-known Breit-Wigner formula\cite{BW36}. The energy-dependence arises just from the first term of the denominator.

For a \emph{many}-level system, such as a real molecule coupled to two baths, using the NEGF approach\cite{TGW01,transiesta} in the linear-response regime and at zero-temperature, the transmission can be expressed as:
\ben
T(E)=\mbox{Tr}\left\{\mg_L(E)\mgf_C(E)\mg_R(E)\mgf_C(E)^\dagger\right\},
\label{land}
\een
where the bold symbols are matrices of dimension of the subspace relevant to the ``extended molecule'' region, usually the molecule plus additional lead layers on either side. $\mgf_C(E)=\left[E\ms_C-\mh_C-\msig_L(E)-\msig_R(E)\right]^{-1}$ is the Green's function of the extended molecule, and $\mg_L(E)=i\left[\msig_L(E)-\msig_L(E)^\dagger\right]$ and similarly for $\mg_R(E)$. $\msig_L(E)$ and $\msig_R(E)$ are the self-energies due to the left and right lead, respectively. In \emph{ab initio} calculations, $\mh_C$, the Hamiltonian of the central region, is usually approximated\cite{TGW01,transiesta} by the Kohn-Sham Hamiltonian $\mh_{\rm S}$ of density functional theory (DFT), or $\mh_{\rm S}+\msig_{\rm mol}$ in \dfts\cite{QVCL07,LWYA14}.

\begin{figure}
\begin{center}
\includegraphics[width=3.5in]{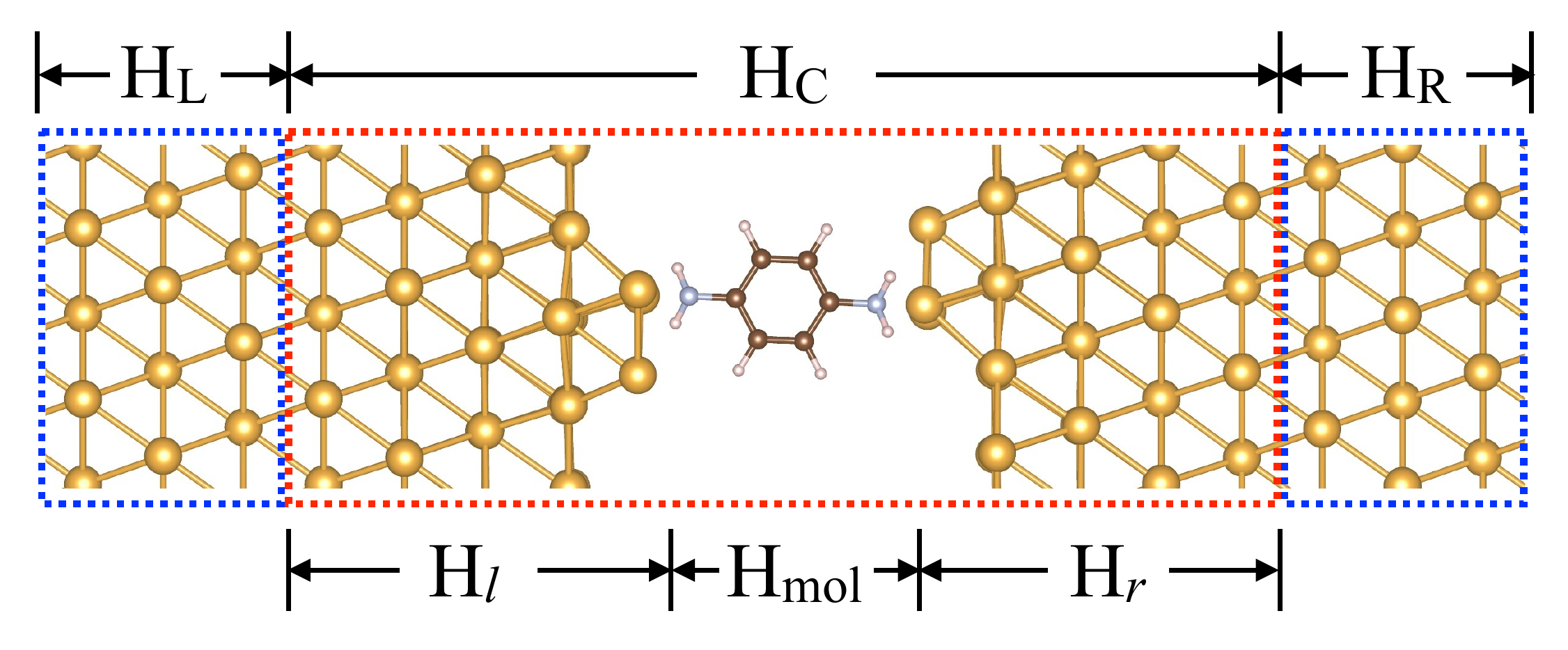}
\caption{A typical system in DFT-NEGF calculation, Au-benzenediamine(BDA)-Au junction, used to define the regions discussed in the text. The molecule is generic and is not limited to the BDA shown in the figure. The red box indicates the extended molecule, consisting of the molecule and several layers of lead atoms on the left and on the right. The blue boxes are left and right leads, which extend to $-\infty$ and $+\infty$, respectively.}
\label{f:cart}
\end{center}
\end{figure}

Fig. \ref{f:cart} shows a typical system considered in DFT-NEGF calculations. The molecule and several layers of leads on the left and right in the red box make up the extended molecule. Using an atom-centered basis, if we denote the dimension of the basis set of the extended molecule as $N_C$, then the matrices in Eq. \eqref{land} are of size $N_C\times N_C$. Using an extended molecule rather than the bare molecule in the scattering central region captures important screening and chemical effects associated with the leads\cite{DPL00b}, but it is difficult to understand the shape of $T(E)$ in terms of molecular resonances \cite{SGPF06}, because the wavefunctions of the extended molecule are combinations of both molecular and lead orbitals. An eigenchannel analysis \cite{PB07} can help understand transport in terms of the \emph{extended} molecule, but it is often challenging to interpret the eigenchannels in terms of contributions of \emph{bare} molecular orbitals \cite{CLM98,HCWS02,LHZQ06}. Throughout this paper, by ``bare molecular orbitals'' we mean the eigenvectors of the molecular subblock of the Hamiltonian in Eq. \eqref{mhc} below.

In order to understand $T(E)$ in terms of resonance states originating from bare molecular orbitals, we re-express Eq. \eqref{land} in terms of matrices of size $N_m \times N_m$, where $N_m$ is system size of the molecule and $N_m < N_C$. We write $\mh_C$ as
\ben
\mh_C=\begin{pmatrix}
\mh_l & \mv_{lm} & \mv_{lr} \\
\mv_{ml} & \mh_{\rm mol} & \mv_{mr} \\
\mv_{rl} & \mv_{rm} & \mh_r \\
\end{pmatrix}.
\label{mhc}
\een
A similar expression exists for the overlap matrix $\ms_C$. Here, the subscript ``mol'' denotes the bare molecule, and $l$ and $r$ the lead atoms in the ``extended molecule''. 

In what follows, we first neglect the direct coupling between the left and right lead atoms in the central region, $\mv_{lr}$ and $\mv_{rl}$ (and corresponding matrix subblocks in $\ms_C$), as in Ref. \cite{JP06}. We then express $\mathbf{G}_C$ in a similar $3 \times 3$ block-form; and, using $(E\ms_C-\mh_C-\msig_L-\msig_R)^{-1}\mgf_C=\mathbf{I}$ and the fact that $\msig_L$ is non-zero only in the upper left corner ($l$ part) and $\msig_R$ is non-zero only in the lower right corner ($r$ part)\cite{B14t}, we can solve for $\mgf_{\rm mol}$, the central block in $\mgf_C$, and reexpress the transmission as
\ben
T(E)=\mbox{Tr}\left\{\tmg_L(E)\mgf_{\rm mol}(E)\tmg_R(E)\mgf_{\rm mol}(E)^\dagger\right\},
\label{landmol}
\een
where all quantities in bold are matrices of the size $N_m \times N_m$, with
\ben
\mgf_{\rm mol}=\left[E\ms-\mh_{\rm mol}-\tmsig_L(E)-\tmsig_R(E)\right]^{-1},
\label{gfmol}
\een
and
\ben
\tmsig_L(E)=\left[E\ms_{ml}-\mv_{ml}\right]\left[E\ms_l-\mh_l-\msig_L(E)\right]^{-1}\left[E\ms_{lm}-\mv_{lm}\right].
\label{tmsigl}
\een
In Eq. \eqref{landmol}, $\tmg_L(E)=i\left[\tmsig_L(E)-\tmsig_L(E)^\dagger\right]$, with similar expression for $\tmsig_R(E)$ and $\tmg_R(E)$ \cite{JP06}. Eq. \eqref{landmol} is formally analogous to Eq. \eqref{land}, but is of dimension $N_m \times N_m$ rather than $N_C \times N_C$.

We note that in Eq. \eqref{landmol}, if $\tmg_L(E)$, $\tmg_R(E)$, and $\mgf_{\rm mol}(E)$ are simultaneously diagonalizable at all energies, then $T(E)$ can be expressed as a sum of Lorentzian-like [Eq. \eqref{lore}] terms, with energy-dependent broadening functions. However this is not the case in general. We note that in past work\cite{SGPF06}, $\mgf(E)$ and $\mgf(E)^\dagger$ were diagonalized simultaneously in Eq. \eqref{land} and $\mg_L(E)$ and $\mg_R(E)$ were expressed in the eigenbasis of $\mgf(E)$. 

Here, instead of diagonalizing the Green's function, we diagonalize the following four matrices:
\ben
\mh_{\rm mol}\left.\left|\psi_j\right>\right.=\epsilon_j\left.\left|\psi_j\right>\right.;
\label{diagmol}
\een
\ben
\left[\mh_{\rm mol}+\tmsig_L\right]\left.\left|\psi^L_i\right>\right.=\left[\epsilon^L_i-i\Gamma^L_i/2\right]\left.\left|\psi^L_i\right>\right.;
\label{diagl}
\een
\ben
\left[\mh_{\rm mol}+\tmsig_R\right]\left.\left|\psi^R_i\right>\right.=\left[\epsilon^R_i-i\Gamma^R_i/2\right]\left.\left|\psi^R_i\right>\right.;
\label{diagr}
\een
and
\ben
\left[\mh_{\rm mol}+\tmsig_T\right]\left.\left|\psi^T_i\right>\right.=\left[\epsilon^T_i-i\Gamma^T_i/2\right]\left.\left|\psi^T_i\right>\right..
\label{diagt}
\een
In Eq. \eqref{diagt}, $\tmsig_T=\tmsig_L+\tmsig_R$. In Eqs. \eqref{diagl}-\eqref{diagt}, all the quantities are explicitly energy-dependent except for $\mh_{\rm mol}$, and for simplicity we suppress the energy dependence. The matrices in Eqs. \eqref{diagl}-\eqref{diagt} are non-Hermitian, and the eigenvalues are therefore complex. We note in passing that because of the nonorthogonal basis and the overlap matrix $\ms$, Eqs. \eqref{diagmol}-\eqref{diagt} correspond to generalized eigenvalue problems. The eigenvectors in Eq. \eqref{diagmol} will be very similar to isolated gas phase molecular orbitals, provided that the molecule is weakly-coupled to leads. The eigenvectors in Eqs. \eqref{diagl}-\eqref{diagt} are resonance states originating from bare molecular orbitals $\left.\left|\psi_j\right>\right.$ in Eq. \eqref{diagmol}, and the imaginary parts of the eigenvalues can be interpreted as resonance widths \cite{Datta05,SYG13}.

For each molecular orbital $\left.\left|\psi_j\right>\right.$ of interest in Eq. \eqref{diagmol}, we need to identify corresponding eigenstates in Eqs. \eqref{diagl}-\eqref{diagt}. For example, if, as in many systems, the conductance is dominated by a nearby HOMO resonance, then we can simply take $j$=HOMO in Eq. \eqref{diagmol}. We compute the projection $\left|\left<\psi_j\left|\psi^L_i\right.\right>\right|^2$ for each $\left.\left|\psi^L_i\right.\right>$, and assign the imaginary part (multiplied by $-2$) of the eigenvalue corresponding to the largest projection as resonance width $\Gamma_j$ for $j$-th molecular orbital $\left.\left|\psi_j\right>\right.$ in Eq. \eqref{diagmol}. This procedure is carried out at every energy $E$, and the resulting $\Gamma^L_j(E)$ is energy-dependent. For levels for which the largest projection is nearly unity ($\sim0.99$), the energy-dependent width may be meaningfully assigned to a molecular orbital. 

We can apply this strategy to $\left.\left|\psi^R_i\right>\right.$ and $\left.\left|\psi^T_i\right>\right.$ in Eqs. \eqref{diagr} and \eqref{diagt} and similarly define $\Gamma^R_j(E)$ and $\Gamma^T_j(E)$, respectively. Strictly speaking, $\Gamma^T_j(E)$ is not necessarily equal to $\Gamma^L_j(E)+\Gamma^R_j(E)$, but for molecular orbitals whose largest projection are nearly unity, the difference will be negligible. In Eq. \eqref{diagt}, $\epsilon^T_i(E)$ corresponding to the largest projection is the resonance energy for molecular orbital $\left.\left|\psi_j\right>\right.$, different from $\epsilon_j$ in Eq. \eqref{diagmol}.

In the energy range of interest, usually within a few eV around the resonance, $T(E)$ is reproduced using the energy-dependent broadening functions defined above in a Lorentzian-like formula:
\ben
T(E)\approx\sum_{j,\epsilon_j\in \Delta E}\frac{\Gamma^L_j(E)\Gamma^R_j(E)}{\left[E-\epsilon^T_j(E)\right]^2+\left[\Gamma^L_j(E)+\Gamma^R_j(E)\right]^2/4},
\label{landlore}
\een
where $\Delta E$ is some pre-defined energy range. One can compare Eq. \eqref{landlore} with Eq. \eqref{lore} and they are very similar from a formal point of view. However they are conceptually different: Eq. \eqref{lore} only applies rigorously to a one-level model system or many non-interacting levels, but Eq. \eqref{landlore} is a good  approximation to the more complicated, interacting many-level systems such as realistic molecular junctions, and becomes exact when Eqs. \eqref{diagl}-\eqref{diagt} are simultaneously diagonalizable. In practice, periodic boundary conditions are used along directions transverse to current flow, and Eq. \eqref{landlore} is weighted over $\{k_\parallel\}$.

\begin{figure}
\begin{center}
\includegraphics[width=3.5in]{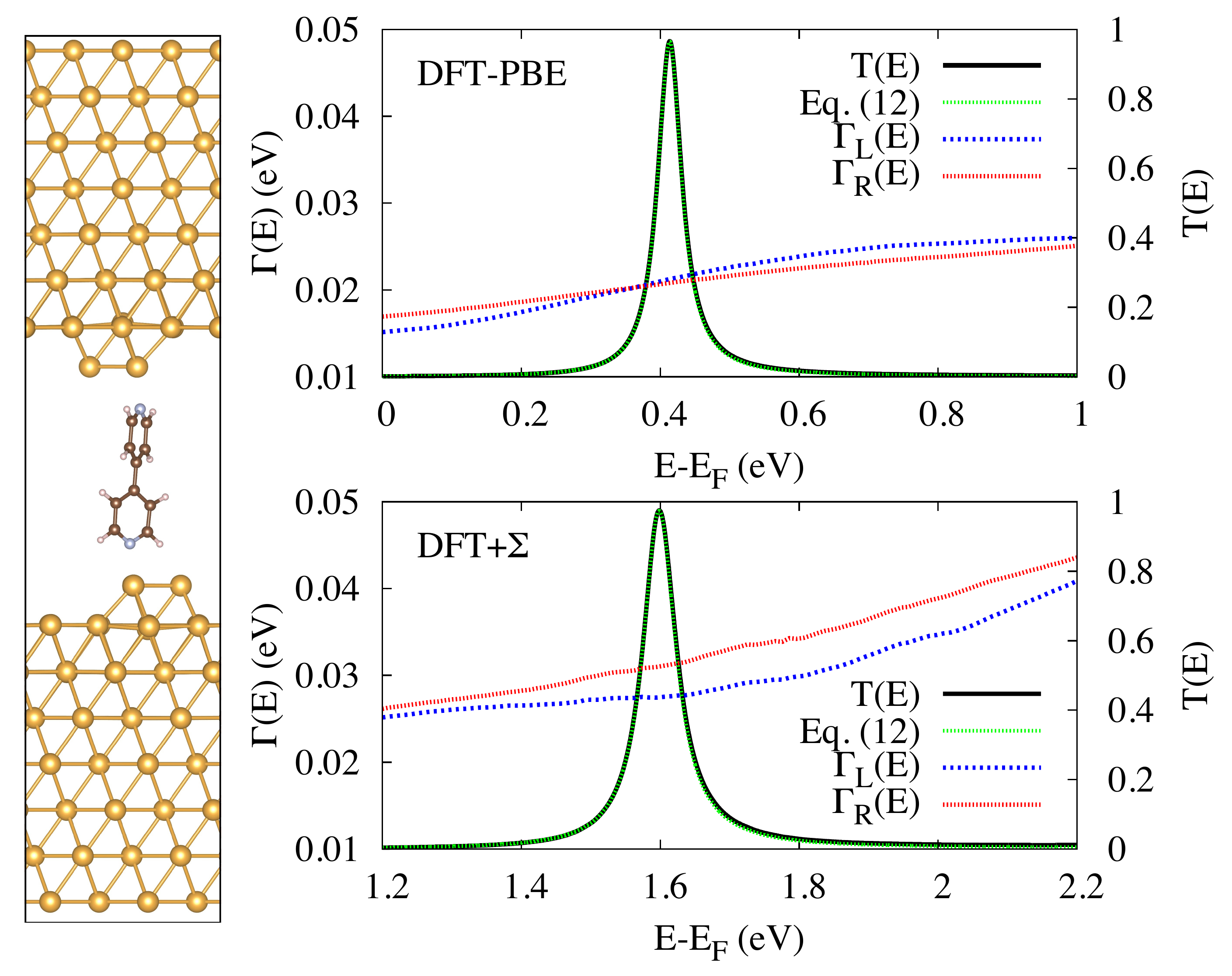}
\caption{Left: Au-Bipyridine-Au junction structure from Ref. \cite{DWCV12}. Right: $T(E)$ and $k_\parallel$-averaged broadening functions $\Gamma_L(E)$ and $\Gamma_R(E)$ for the LUMO resonance, calculated from DFT-PBE (upper panel) and \dfts (lower panel). $T(E)$ is well reproduced by the energy-dependent broadening functions using Eq. \eqref{landlore}.}
\label{f:bp}
\end{center}
\end{figure}

Having described this new approach above, we now turn to the discussion of its implementation and application. We implement the method in the TranSIESTA \cite{transiesta} package, which is based on the NEGF framework. We first apply the method to a weakly-coupled system, an Au-Bipyridine-Au junction. In this system, the LUMO is the conducting orbital\cite{DWCV12}, and we focus on the energy range around the LUMO resonance. Fig. \ref{f:bp} shows the structure of the junction, and upper panel shows the $T(E)$ curve calculated from DFT-NEGF implemented in TranSIESTA \cite{transiesta} using the PBE functional\cite{PBE96}. A $16\times 16$ $k_\parallel$-mesh is used in the calculation of $T(E)$. $T(E_F)$ is clearly dominated by LUMO resonance around 0.4 eV above Fermi level. Also shown are the $k_\parallel$-averaged $\Gamma_L(E)$ and $\Gamma_R(E)$ for the LUMO resonance: $\Gamma_L(E)=\sum_{k,k\in\left\{k_\parallel\right\}}w_k\Gamma_L(k;E)$, where $w_k$ is the weight of a $k_\parallel$ point, and similarly for $\Gamma_R(E)$. To demonstrate that the energy-dependent LUMO broadening functions reproduce $T(E)$, we calculate $T(E)$ based on Eq. \eqref{landlore} ($j$=LUMO) and average over $\{k_\parallel\}$. It matches the NEGF $T(E)$ very well. In Fig. \ref{f:bp}, $\Gamma_L(E)$ and $\Gamma_R(E)$ are not identical, because the junction is not entirely symmetric. We note that in symmetric junctions, $\Gamma_L(E)=\Gamma_R(E)$ for any symmetric molecular orbital at any energy. 

It is known that conventional local or semi-local functionals, such as PBE, overestimate the linear-response conductance by an order of magnitude or more. This is due to its underestimation of level alignment between junction Fermi energy and conducting orbital\cite{NHL06}. Beyond DFT, GW-based approaches\cite{QVCL07,TR08,LWYA14} can correct level alignment, leading to quantitative agreement with experiment in some cases. In Ref.\cite{LWYA14}, we showed that using one such approach, \dfts\cite{QVCL07,LWYA14}, the $\mgf(E)$ of Eq. \eqref{land} is modified by replacing $\mh_{\rm mol}$ from $\mh_{\rm S}$ to $\mh_{\rm S}+\msig_{\rm mol}$, i.e., a $\msig_{\rm mol}$ correction to the PBE Kohn-Sham Hamiltonian. Although \dfts does not modify $\mg_L(E)$ and $\mg_R(E)$ matrices in Eq. \eqref{land}, it changes the broadening of molecular resonance implicitly, as can be well understood using the method developed in this work. In Eqs. \eqref{diagl}-\eqref{diagt}, the $\tmsig(E)$ is not altered in \dfts from PBE, but $\mh_{\rm mol}$ is. As a result, the eigenvalues, and in particular their imaginary parts, also change.

To show this explicitly, we carry out \dfts calculations of the same Au-Bipyridine-Au junction, with a shift of PBE unoccupied orbital energies upward by 1.2 eV\cite{DWCV12} and PBE occupied orbital energies downward by the same amount. In lower panel of Fig. \ref{f:bp}, we show the \dfts results for $\Gamma_L(E)$ and $\Gamma_R(E)$ of the LUMO resonance, which are slightly larger than their PBE counterparts, leading to a broader resonance.

\begin{figure}
\begin{center}
\includegraphics[width=3.5in]{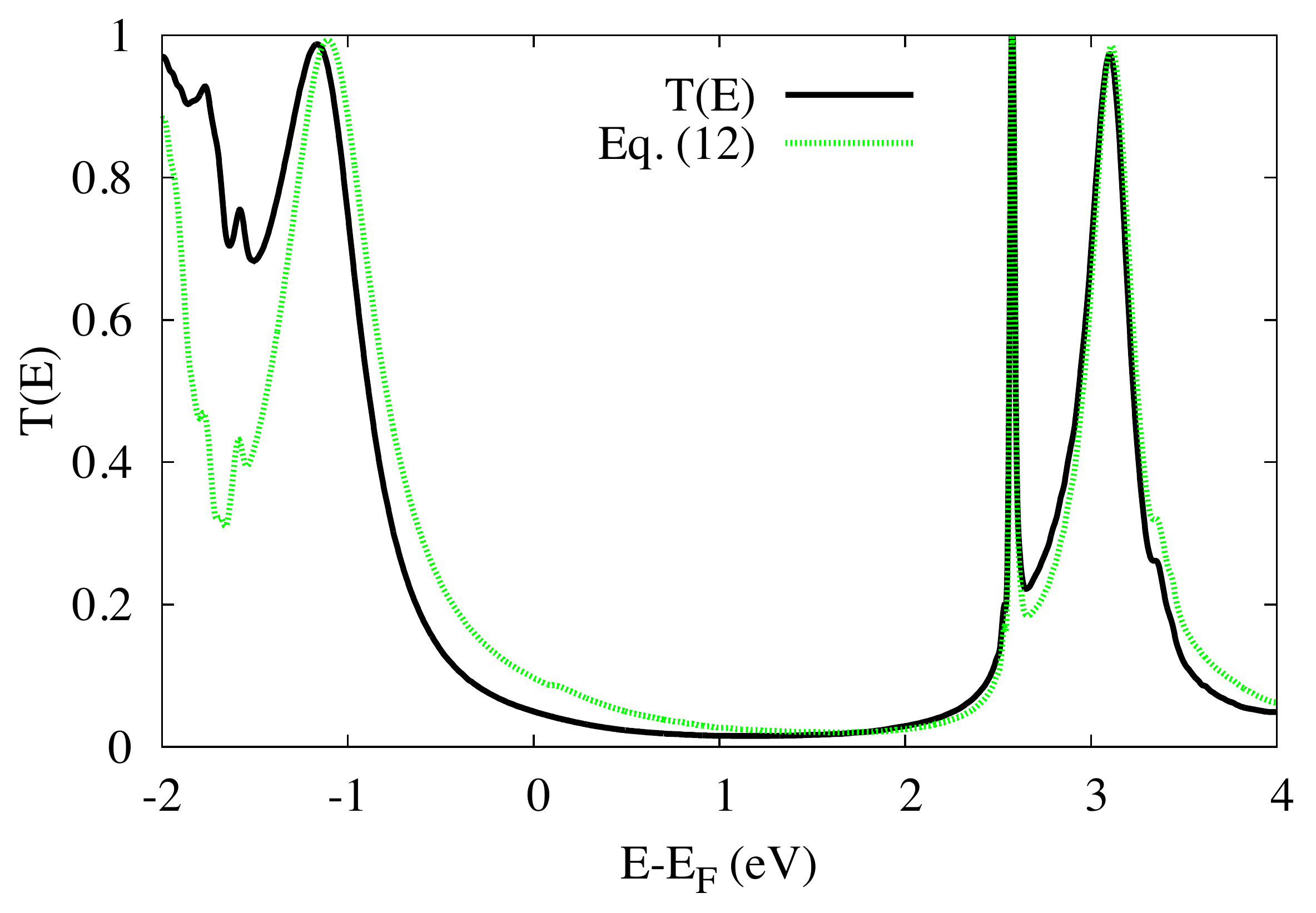}
\caption{$T(E)$ of a Au-BDA-Au junction, calculated both from DFT-NEGF and from Eq. \eqref{landlore}. For LUMO and LUMO+1 resonances, Eq. \eqref{landlore} reproduces exact result well.}
\label{f:bda}
\end{center}
\end{figure}

In the previous example, the LUMO dominates conductance, and in Eq. \eqref{landlore}, we simply take $j$=LUMO. In general, we can extend the sum in Eq. \eqref{landlore} over a few molecular orbitals near the junction Fermi level. We show in Fig. \ref{f:bda} the $T(E)$ calculated from Eq. \eqref{landlore} for Au-BDA-Au junction of Fig. \ref{f:cart}. In this sytem, we sum over $j$=HOMO, LUMO, and LUMO+1. We can see that for LUMO and LUMO+1, Eq. \eqref{landlore} reproduces the exact result very well, since $\left|\left<\psi_j\left|\psi^{L,R}_i\right.\right>\right|^2 \sim 1$ in both cases. For the HOMO resonance, Eq. \eqref{landlore} is not quantitative due to more significant hybridization between the HOMO and lead states. Indeed, in this case $\left|\left<\psi_{\rm HOMO}\left|\psi^{L,R}_i\right.\right>\right|^2 \sim 0.8$, and attributing a single orbital to this resonance is less justified.

\begin{figure}
\begin{center}
\includegraphics[width=3.5in]{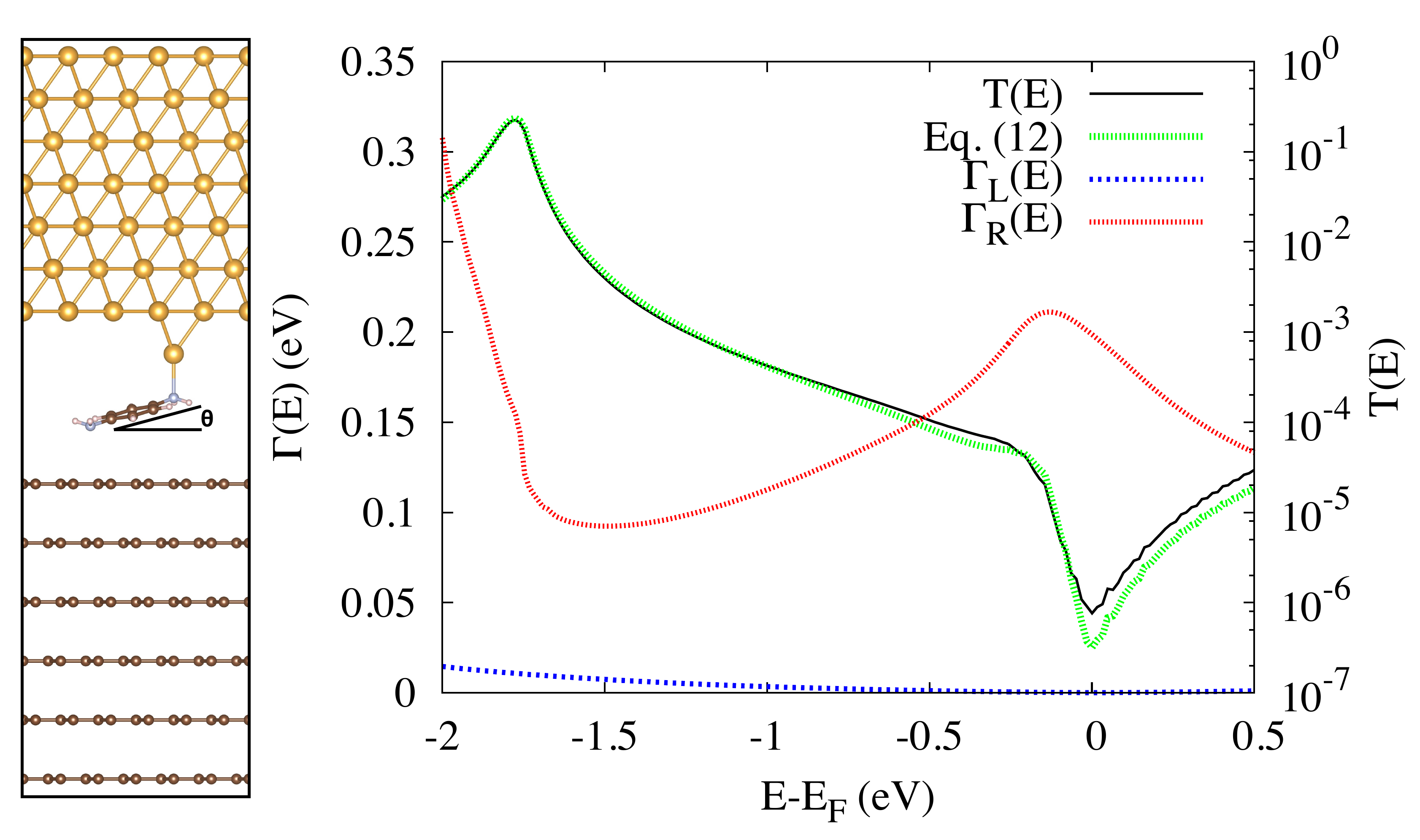}
\caption{Au-benzenediamine-graphite junction ($\theta=14^{\circ}$) and its \dfts $T(E)$ curve \cite{KLLN14}. The broadening functions for the HOMO resonance due to graphite $[\Gamma_L(E)]$ and gold $[\Gamma_R(E)]$ are shown. Eq. \eqref{landlore} faithfully reproduces $T(E)$ calculated from NEGF even when wide band limit breaks down.}
\label{f:gra}
\end{center}
\end{figure}

Additional advantages of the method can be demonstrated for asymmetric junctions, such as Au-benzenediamine-graphite\cite{KLLN14}, where $\Gamma$ can be conveniently decomposed into $\Gamma_L$ and $\Gamma_R$. Fig. \ref{f:gra} shows the structure of the junction, and its \dfts $T(E)$ curve, as well as the broadening functions $\Gamma_L(E)$ and $\Gamma_R(E)$ for HOMO resonance. Details of the system setup and calculations can be found in Ref.\cite{KLLN14}. The transmission is $\theta$-dependent, where $\theta$ is the angle between the molecule and graphite surface. In Fig. \ref{f:gra}, $\theta=14^{\circ}$, the left electrode is graphite, and the right electrode is gold. $\Gamma_L(E)$ is much smaller than $\Gamma_R(E)$, which leads to a $T(E)$ peak value significantly less than unity. Additionally, a simple one-level, wide-band-limit-based Lorentzian model [Eq. \eqref{wbl}] can not describe $T(E)$ in this case, because the graphite density of states around Fermi level is highly energy-dependent and the wide band limit breaks down. However, with Eq. \eqref{landlore}, a Lorentizian-like equation with energy-dependent broadening functions for HOMO resonance well captures and explains the features of $T(E)$ curve.

\begin{figure}
\begin{center}
\includegraphics[width=3.5in]{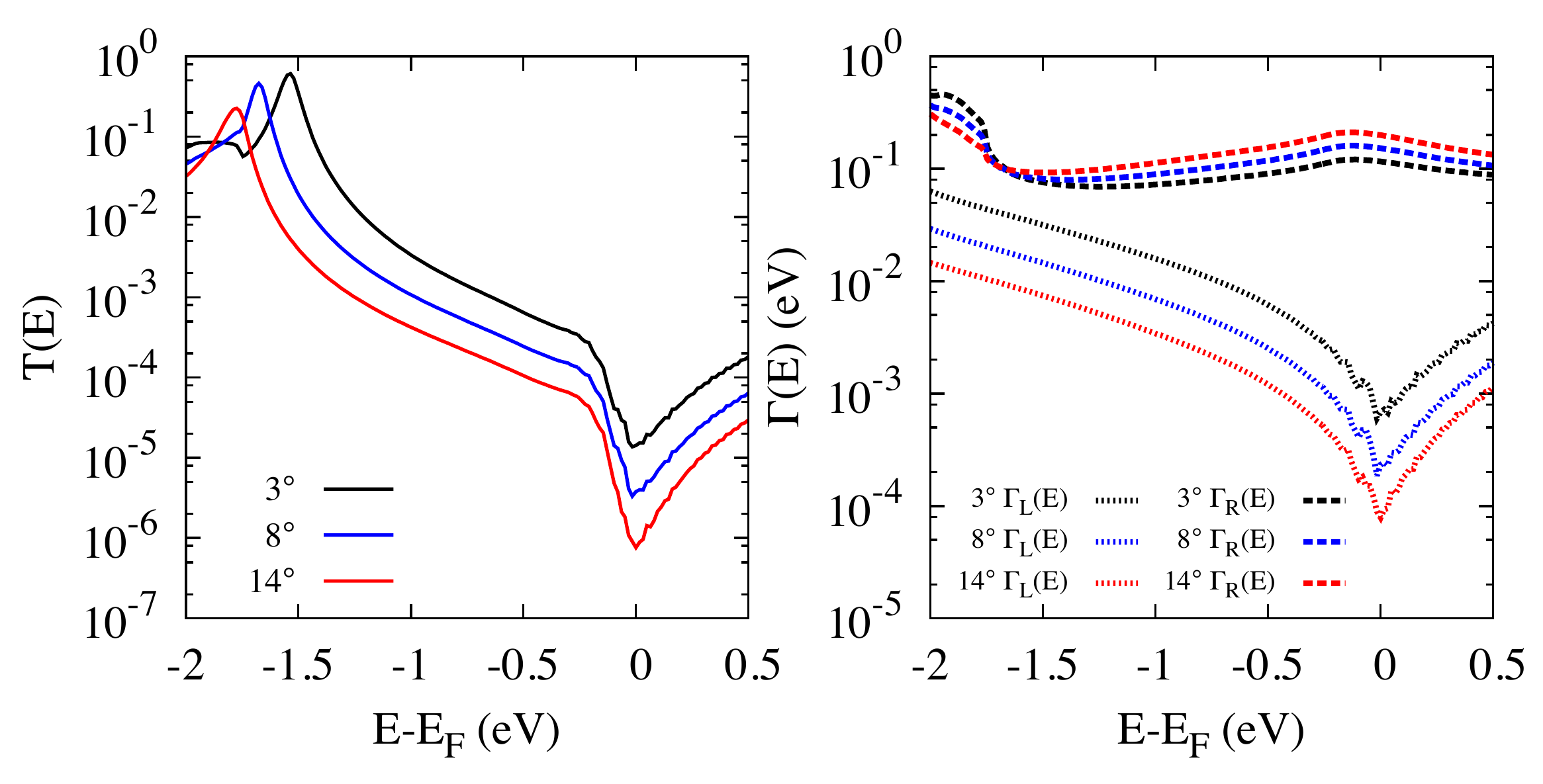}
\caption{Left panel: Computed angle-dependence $T(E)$ for the Au-benzenediamine-graphite junction of Fig. \ref{f:gra}. Right panel: Energy-dependent broadening functions $\Gamma_L(E)$ and $\Gamma_R(E)$ for the HOMO resonance. The left electrode is graphite, and the right electrode is gold.}
\label{f:gragamma}
\end{center}
\end{figure}

In Ref.\cite{KLLN14}, a $\theta$-dependent $T(E)$ is calculated, and the $T(E)$ peak at the HOMO resonance becomes smaller as the angle increases, which implies that the coupling becomes more asymmetric with angle. We apply the method developed in this work to three different angles ($3^{\circ},8^{\circ}$, and $14^{\circ}$), and show the results in log-scale in Fig. \ref{f:gragamma}. It is clear that $\Gamma_L(E)$, the coupling of HOMO to graphite, changes dramatically with angle, while $\Gamma_R(E)$, the coupling to gold, is relatively unchanged. This implies that the wide band limit breaks down completely for graphite electrode, especially around Fermi level, but that it holds for gold electrode, following conventional intuition\cite{VST13}. 

In conclusion, in this work we develop a new approach to compute energy-dependent resonance broadening in both symmetric and asymmetric molecular junctions, based on non-equilibrium Green's function formalism. The method is based on two steps: (1) re-expressing Landauer formula in terms of quantities of \emph{extended} molecule [Eq. \eqref{land}] to a formula in terms of quantities of \emph{bare} molecule [Eq. \eqref{landmol}]; and (2) diagonalization of non-Hermitian matrices and recognition of imaginary part of eigenvalues as broadening functions. These energy-dependent broadening functions, together with a Lorentzian-like formula, Eq. \eqref{landlore}, reproduce the $T(E)$ calculated from NEGF very well. The method is then applied to different molecular junctions, both symmetric and asymmetric, and in the latter case, resonance broadening can be decomposed into components due to the left lead and the right lead, respectively. The method also explains clearly why \dfts changes the phenomenological broadening of molecular resonances implicitly, without changing the $\mg_L$ and $\mg_R$ matrices in the Landauer formula. 

The method developed in this work is not limited to molecular junctions, but can be applied to study any resonance states provided that a suitable Green's function treatment exists. For example, it can be applied to adsorbate molecules on surfaces, where the molecular levels are broadened, as evidenced in the spectral function, due to coupling to the semi-infinite bulk material. Standard slab DFT calculations use a \emph{finite} number of layers for the surface and a large vacuum region with periodic boundary conditions. These calculations can yield projected density of states of the molecule, but with artificial broadening. With the NEGF formalism and the method developed in this work, it is possible to treat the effect of \emph{semi-infinite} bulk materials as a self-energy and compute the molecular resonance broadening on a surface. To be specific, Eqs. \eqref{gfmol}-\eqref{diagl} and the definition of energy-dependent $\Gamma(E)$ still apply, but with only one lead. This is an interesting area of future study.

Z.-F. L. thanks Yu Zhang for helpful discussions. This work is supported by the U.S. Department of Energy, Office of Basic Energy Sciences, Materials Sciences and Engineering Division, under Contract No. DE-AC02-05CH11231. This work is also supported by the Molecular Foundry through the U.S. Department of Energy, Office of Basic Energy Sciences under the same contract number. Portion of the computation work is done using NERSC resources.
\bibliography{lit_broadening}
\end{document}